\documentclass[preprint,prd,aps,showpacs,showkeys,nofootinbib]{revtex4}
\usepackage{amssymb}
\usepackage{}
\usepackage{amsmath}
\allowdisplaybreaks[4]
\usepackage{multirow}
\usepackage{graphicx}
\usepackage{float}

\begin{document}

\preprint{}

\title{Muon conversion to electron in nuclei in Minimal R-symmetric Supersymmetric Standard Model}

\author{Ke-Sheng Sun$^a$\footnote{sunkesheng@126.com, sunkesheng@bdu.edu.cn}, Sheng-Kai Cui$^{b,c}$\footnote{2252953633@qq.com}, Wei Li$^{b,c}$\footnote{watliwei@163.com}, Hai-Bin Zhang$^{b,c}$\footnote{hbzhang@hbu.edu.cn}}

\affiliation{$^a$Department of Physics, Baoding University, Baoding 071000,China\\
$^b$Department of Physics, Hebei University, Baoding 071002, China\\
$^c$Key Laboratory of High-precision Computation and Application of
Quantum Field Theory of Hebei Province, Baoding, 071002, China}

\begin{abstract}

We analyze the lepton flavor violating process $\mu$-e conversion in the framework of the minimal R-symmetric supersymmetric standard model. The theoretical predictions are determined by considering the experimental constraint on parameter $\delta^{12}$ from the lepton flavor violating decay $\mu\rightarrow e \gamma$. The predictions for CR($\mu-e$,Nucleus) in nuclei are not sensitive to tan$\beta$ or $m_A$ and take values in a narrow region. The numerical results show that $\gamma$ penguins dominate the predictions on CR($\mu-e$,Nucleus), and the contribution from Higgs penguins is insignificant. The Z penguins and box diagrams are less dominant in the predictions on CR($\mu-e$,Nucleus) in a large parameter region. For small squark mass parameter, the contribution from box diagrams is comparable with dipole contribution from $\gamma$ penguins. The theoretical predictions on conversion rate CR($\mu-e$,Nucleus) in a Al or Ti target can be enhanced close to the future experimental sensitivities and are very promising to be observed in near future experiment.

\end{abstract}



\maketitle

\section{Introduction\label{sec1}}
\indent\indent

Searching for Lepton Flavor Violating (LFV) decays are of great importance in probing New Physics (NP) beyond the Standard Model (SM) in which the theoretical predictions on those LFV decays are suppressed by small masses of neutrinos and far beyond the experimental accessibility. There are many different ways to search LFV such as $\mu\rightarrow e\gamma$, $\mu\rightarrow 3e$, $\mu-e$ conversion in nucleus, $\tau$ decays, hadron decays and so on. However, no LFV signals have been observed in experiment up to now. The $\mu-e$ conversion in nucleus is a process that muons are captured in a target of atomic nucleus and form a muonic atom. Several experiments have been built or planned to built to search for this process. Current limit on the $\mu-e$ conversion rate is $4.6\times 10^{-12}$ for a Ti target at TRIUMF \cite{TRIUMF}, $4.3\times 10^{-12}$ for a Ti target and $7\times 10^{-13}$ for a Au target at SINDRUM-II experiment \cite{SINDRUMII}. In future, this LFV process may be observed by experiments with improved sensitivity. A future prospects of $10^{-13}$ for a C target or $10^{-14}$ for a SiC target at DeeMe \cite{DeeMe}, $10^{-18}$ for a Ti target at PRISM \cite{PRISM} and $10^{-16}-10^{-17}$ for a Al target at Mu2e and COMET \cite{Mu2e, COMET} will be achieved, which improve the current experimental limits by several orders of magnitude.

The $\mu-e$ conversion rate has been calculated in the literature for various extensions of SM. Some seesaw models with right handed neutrinos \cite{Riazuddin, Chang, Ioannisian, Pilaftsis, Deppisch, Ilakovac, Deppisch2}, scalar triplets \cite{Raidal, Ma, Dinh}, fermion singlet \cite{sksis} and fermion triplets \cite{Abada}, can have CR$(\mu-e,Nucleus)$ close to the experimental sensitivity.
There are a few studies within models of non-SUSY, such as unparticle model \cite{sksup,Ding}, littlest Higgs model \cite{Blanke,Aguila}, left-right symmetric models \cite{Bonilla}, 331 model \cite{Huong} and so on. There are also a few studies within models of SUSY, such as MSSM \cite{Hisano}, R-parity violating SUSY \cite{Sato}, low-scale seesaw models of minimal supergravity \cite{Ilakovac2013}, BLMSSM \cite{Guo, Dong}, the CMSSM-seesaw \cite{Arganda}, $\mu\nu$SSM \cite{Zhang} and so on. $\mu-e$ conversion is particularly sensitive to Higgs mediated LFV because it is not suppressed by small Yukawa couplings as $\mu\rightarrow e\gamma$ and $\mu\rightarrow 3e$, and Higgs-induced LFV occurs in many NP models \cite{Crivellin}.
Some pedagogical introductions on the theoretical motivations for charged LFV and the experimental aspects is provided in Ref. \cite{Calibbi,Bernstein,Lindner}.

In this paper, we will study the LFV process $\mu-e$ conversion in the Minimal R-symmetric Supersymmetric Standard Model (MRSSM) \cite{Kribs}. The MRSSM has an unbroken global $U(1)_R$ symmetry and provides a new solution to the supersymmetric flavor problem in MSSM. In this model, R-symmetry forbids Majorana gaugino masses, $\mu$ term, $A$ terms and all left-right squark and slepton mass mixings. The $R$-charged Higgs $SU(2)_L$ doublets $\hat{R}_u$ and $\hat{R}_d$ are introduced in MRSSM to yield the Dirac mass terms of higgsinos.	Additional superfields $\hat{S}$, $\hat{T}$ and $\hat{O}$ are introduced to yield Dirac mass terms of gauginos. Studies on phenomenology in MRSSM can be found in literatures \cite{Die1, Die2, Die3, Die4, Die5,Die6,KSS, Kumar, Blechman, Kribs1, Frugiuele, Jan, Chakraborty, Braathen, Athron, Alvarado,sks1,sks2,sks3}. Similar to MSSM, the off-diagonal entries $\delta^{ij}$ in slepton mass matrices $m_l^2$ and $m_r^2$ dominate the LFV process $\mu-e$ conversion. Taking account of the constraints from radiative decays $\mu\rightarrow e\gamma$ on the off-diagonal parameters $\delta^{ij}$, we explore $\mu-e$ conversion rate as a function of off-diagonal parameter $\delta^{ij}$ and other model parameters.

The paper is organized as follows. In Section \ref{sec2}, we present the details of the MRSSM. All relevant mass matrices and mixing matrices are provided. Feynman diagrams contributing to $\mu-e$ conversion in MRSSM are given at one loop level. The $\mu-e$ conversion rate are computed in effective Lagrangian method, and notations and conventions for effective operators and Wilson coefficients are also listed. The numerical results are presented in Section \ref{sec3}, and the conclusion is drawn in Section \ref{sec4}.

\section{MRSSM\label{sec2}}

In this section, we firstly provide a simple overview of MRSSM in order to fix the notations we use in this paper. The MRSSM has the same gauge symmetry $SU(3)_C\times SU(2)_L\times U(1)_Y$ as the SM and MSSM. The spectrum of fields in MRSSM contains the standard MSSM matter, Higgs and gauge superfields augmented by chiral adjoints $\hat{\cal O},\hat{T},\hat{S}$ and two $R$-Higgs iso-doublets.
The general form of the superpotential of the MRSSM is given by \cite{Die1}
\begin{equation}
\begin{array}{l}
\mathcal{W}_{MRSSM} = \mu_d(\hat{R}_d\hat{H}_d)+\mu_u(\hat{R}_u\hat{H}_u)+\Lambda_d(\hat{R}_d\hat{T})\hat{H}_d+\Lambda_u(\hat{R}_u\hat{T})\hat{H}_u
+\lambda_d\hat{S}(\hat{R}_d\hat{H}_d)\\
\hspace{5.5em}+\lambda_u\hat{S}(\hat{R}_u\hat{H}_u)-Y_d\hat{d}(\hat{q}\hat{H}_d)-Y_e\hat{e}(\hat{l}\hat{H}_d)+Y_u\hat{u}(\hat{q}\hat{H}_u),
\end{array}\label{sptl}
\end{equation}
where $\hat{H}_u$ and $\hat{H}_d$ are the MSSM-like Higgs weak iso-doublets, $\hat{R}_u$ and $\hat{R}_d$ are the $R$-charged Higgs $SU(2)_L$ doublets and the corresponding Dirac higgsino mass parameters are denoted as $\mu_u$ and $\mu_d$. Although R-symmetry forbids the $\mu$ terms of the MSSM, the bilinear combinations of the normal Higgs $SU(2)_L$ doublets $\hat{H}_u$ and $\hat{H}_d$ with the Higgs $SU(2)_L$ doublets $\hat{R}_u$ and $\hat{R}_d$ are allowed in Eq.(\ref{sptl}). Parameters $\lambda_u$, $\lambda_d$, $\Lambda_u$ and $\Lambda_d$ are Yukawa-like trilinear terms involving the singlet $\hat{S}$ and the triplet $\hat{T}$.
For our phenomenological studies we take the soft-breaking terms involving scalar mass that have been considered in \cite{Die3}
\begin{equation}
\begin{array}{l}
V_{SB,S} = m^2_{H_d}(|H^0_d|^2+|H^{-}_d|^2)+m^2_{H_u}(|H^0_u|^2+|H^{+}_u|^2)+(B_{\mu}(H^-_dH^+_u-H^0_dH^0_u)+h.c.)\\
\hspace{3.5em}+m^2_{R_d}(|R^0_d|^2+|R^{+}_d|^2)+m^2_{R_u}(|R^0_u|^2+|R^{-}_u|^2)+m^2_T(|T^0|^2+|T^-|^2+|T^+|^2)\\
\hspace{3.5em}+m^2_S|S|^2+ m^2_O|O^2|+\tilde{d}^*_{L,i} m_{q,{i j}}^{2} \tilde{d}_{L,j} +\tilde{d}^*_{R,i} m_{d,{i j}}^{2} \tilde{d}_{R,j}+\tilde{u}^*_{L,i}  m_{q,{i j}}^{2} \tilde{u}_{L,j}\\
\hspace{3.5em}+\tilde{u}^*_{R,i}  m_{u,{i j}}^{2} \tilde{u}_{R,j}+\tilde{e}^*_{L,i} m_{l,{i j}}^{2} \tilde{e}_{L,j}+\tilde{e}^*_{R,{i}} m_{r,{i j}}^{2} \tilde{e}_{R,{j}} +\tilde{\nu}^*_{L,i} m_{l,{i j}}^{2} \tilde{\nu}_{L,j}.
\end{array}\label{soft}
\end{equation}
All trilinear scalar couplings involving Higgs bosons to squarks and sleptons are forbidden in Eq.(\ref{soft}) cause the sfermions have an R-charge and these terms are non R-invariant, and this relaxes the flavor problem of the MSSM \cite{Kribs}. The Dirac nature is a manifest feature of MRSSM fermions and the soft-breaking Dirac mass terms of the singlet $\hat{S}$, triplet $\hat{T}$ and octet $\hat{O}$ take the form as
\begin{equation}
V_{SB,DG}=M^B_D\tilde{B}\tilde{S}+M^W_D\tilde{W}^a\tilde{T}^a+M^O_D\tilde{g}\tilde{O}+h.c.,
\label{}
\end{equation}
where $\tilde{B}$, $\tilde{W}$ and $\tilde{g}$ are usually MSSM Weyl fermions. R-Higgs bosons do not develop vacuum expectation values since they carry R-charge 2. After electroweak symmetry breaking the singlet and triplet vacuum expectation values effectively modify the $\mu_u$ and $\mu_d$, and the modified $\mu_i$ parameters are given by
\begin{align}
\mu_d^{eff,+}= \frac{1}{2} \Lambda_d v_T  + \frac{1}{\sqrt{2}} \lambda_d v_S  + \mu_d ,\;\;
\mu_u^{eff,-}= -\frac{1}{2} \Lambda_u v_T  + \frac{1}{\sqrt{2}} \lambda_u v_S  + \mu_u.\nonumber
\end{align}
The $v_T$ and $v_S$ are vacuum expectation values of $\hat{T}$ and $\hat{S}$ which carry R-charge zero.

In the weak basis $(\sigma_d,\sigma_u,\sigma_S,\sigma_T)$, the pseudo-scalar Higgs boson mass matrix and the diagonalization procedure are
\begin{eqnarray}
{\cal M}_{A^0} &=& \left(
\begin{array}{cccc}
B_{\mu} \frac{v_u}{v_d} & B_{\mu} & 0 & 0 \\
B_{\mu} &  B_{\mu} \frac{v_d}{v_u} & 0 & 0 \\
0 & 0 & m_S^2+\frac{\lambda_d^2 v_d^2+\lambda_u^2 v_u^2 }{2} & \frac{\lambda_d\Lambda_d v_d^2-\lambda_u\Lambda_u v_u^2}{2 \sqrt{2}} \\
0 & 0 & \frac{\lambda_d\Lambda_d v_d^2-\lambda_u\Lambda_u v_u^2}{2 \sqrt{2}}& m_T^2+ \frac{\Lambda_d^2 v_d^2+\Lambda_u^2 v_u^2 }{4}\\
\end{array}
\right), Z^A {\cal M}_{A^0} (Z^{A})^{\dagger}={\cal M}_{A^0}^{\textup{diag}} .
\end{eqnarray}
In the weak basis $(\phi_d,\phi_u,\phi_S,\phi_T)$, the scalar Higgs boson mass matrix and the diagonalization procedure are
\begin{eqnarray}
{\cal M}_h &=& \left(
\begin{array}{cc}
{\cal M}_{11}&{\cal M}_{21}^T\\
{\cal M}_{21}&{\cal M}_{22}\\
\end{array}
\right), Z^h {\cal M}_{h} (Z^{h})^{\dagger}={\cal M}_{h}^{\textup{diag}},
\end{eqnarray}
where the submatrices ($c_{\beta}=cos\beta$, $s_{\beta}=sin\beta$) are
\begin{eqnarray}
{\cal M}_{11}&=& \left(
\begin{array}{cc}
m_Z^2c^2_{\beta}+m_A^2s^2_{\beta}&-(m_Z^2+m_A^2)s_{\beta}c_{\beta}\\
-(m_Z^2+m_A^2)s_{\beta}c_{\beta}&m_Z^2s^2_{\beta}+m_A^2c^2_{\beta}\\
\end{array}
\right),\nonumber\\
{\cal M}_{21}&=& \left(
\begin{array}{cc}
v_d(\sqrt{2}\lambda_d\mu_d^{eff,+}-g_1M_B^D)&
v_u(\sqrt{2}\lambda_u\mu_u^{eff,-}+g_1M_B^D) \\
v_d(\Lambda_d\mu_d^{eff,+}+g_2M_W^D)& -
v_u(\Lambda_u\mu_u^{eff,1}+g_2M_W^D) \\
\end{array}
\right),\nonumber\\
{\cal M}_{22}&=& \left(
\begin{array}{cc}
4 (M_B^D)^2+m_S^2+\frac{\lambda_d^2 v_d^2+\lambda_u^2 v_u^2}{2} \;
& \frac{\lambda_d \Lambda_d v_d^2-\lambda_u \Lambda_u v_u^2}{2 \sqrt{2}} \\
 \frac{\lambda_d \Lambda_d v_d^2-\lambda_u \Lambda_u v_u^2}{2 \sqrt{2}} \;
 & 4 (M_W^D)^2+m_T^2+\frac{\Lambda_d^2 v_d^2+\Lambda_u^2 v_u^2}{4}\\
\end{array}
\right).\nonumber
\end{eqnarray}

The number of neutralino degrees of freedom in MRSSM is doubled compared to MSSM as the neutralinos are Dirac-type. In the weak basis of four neutral electroweak two-component fermions $\xi_i$=($\tilde{B}$,$\tilde{W}^0$,$\tilde{R}^0_d$,$\tilde{R}^0_u$) with R-charge 1 and four neutral electroweak two-component fermions $\varsigma_i$=($\tilde{S}$,$\tilde{T}^0$,$\tilde{H}^0_d$,$\tilde{H}^0_u$) with R-charge -1, the neutralino mass matrix and the diagonalization procedure are
\begin{eqnarray}
m_{\chi^0} &=& \left(
\begin{array}{cccc}
M^{B}_D &0 &-\frac{1}{2} g_1 v_d  &\frac{1}{2} g_1 v_u \\
0 &M^{W}_D &\frac{1}{2} g_2 v_d  &-\frac{1}{2} g_2 v_u \\
- \frac{1}{\sqrt{2}} \lambda_d v_d  &-\frac{1}{2} \Lambda_d v_d  &-\mu_d^{eff,+}&0\\
\frac{1}{\sqrt{2}} \lambda_u v_u  &-\frac{1}{2} \Lambda_u v_u  &0 &\mu_u^{eff,-}\end{array}
\right),(N^{1})^{\ast} m_{\chi^0} (N^{2})^{\dagger} =m_{\chi^0}^{\textup{diag}}.
\end{eqnarray}
The mass eigenstates $\kappa_i$ and $\varphi_i$, and physical four-component Dirac neutralinos are
\begin{equation}
\xi_i=\sum^4_{j=1}(N^1_{ji})^{\ast}\kappa_j, \varsigma_i=\sum^4_{j=1}(N^2_{ij})^{\ast}\varphi_j, \chi^0_i=\left(
\begin{array}{c}
\kappa_i\\
\varphi_i^{\ast}\\
\end{array}
\right).\nonumber
\end{equation}

The number of chargino degrees of freedom in MRSSM is also doubled compared to MSSM and these charginos can be grouped to two separated chargino sectors according to their R-charge. The $\chi^{\pm}$-charginos sector has R-charge 1 electric charge; the $\rho$-charginos sector has R-charge -1 electric charge. In the basis $\xi^+_i$=($\tilde{W}^+$, $\tilde{R}^+_d$) and $\varsigma^-_i$=($\tilde{T}^-$, $\tilde{H}^-_d$), the $\chi^{\pm}$-charginos mass matrix and the diagonalization procedure are
\begin{equation}
m_{\chi^{\pm}} = \left(
\begin{array}{cc}
g_2 v_T  + M^{W}_D &\frac{1}{\sqrt{2}} \Lambda_d v_d \\
\frac{1}{\sqrt{2}} g_2 v_d  &-\frac{1}{2} \Lambda_d v_T  + \frac{1}{\sqrt{2}} \lambda_d v_S  + \mu_d\end{array}
\right),(U^{1})^{\ast} m_{\chi^{\pm}} (V^{1})^{\dagger} =m_{\chi^{\pm}}^{\textup{diag}}.
\end{equation}
The mass eigenstates $\lambda^{\pm}_i$ and physical four-component Dirac charginos are
\begin{equation}
\xi^+_i=\sum^2_{j=1}(V^1_{ij})^{\ast}\lambda^+_j, \varsigma^-_i=\sum^2_{j=1}(U^1_{ji})^{\ast}\lambda^-_j, \chi^{\pm}_i=\left(
\begin{array}{c}
\lambda^+_i\\
\lambda_i^{-\ast}\\
\end{array}
\right).\nonumber
\end{equation}
Here, we don't discuss the $\rho$-charginos sector in detail since it doesn't contribute to $\mu-e$ conversion. More information about the $\rho$-charginos can be found in Ref.\cite{Die3,Die5,sks1,KSS}.

In MRSSM the LFV decays mainly originate from the potential misalignment in sleptons mass matrices. In the gauge eigenstate basis $\tilde{\nu}_{iL}$, the sneutrino mass matrix and the diagonalization procedure are
\begin{equation}
m^2_{\tilde{\nu}} =
m_l^2+\frac{1}{8}(g_1^2+g_2^2)( v_{d}^{2}- v_{u}^{2})+g_2 v_T M^{W}_D-g_1 v_S M^{B}_D,
Z^V m^2_{\tilde{\nu}} (Z^{V})^{\dagger} = m^{2,\textup{diag}}_{\tilde{\nu}},\label{sn}
\end{equation}
where the last two terms in mass matrix are newly introduced by MRSSM. The slepton mass matrix and the diagonalization procedure are
\begin{equation}
\begin{array}{l}
m^2_{\tilde{e}} = \left(
\begin{array}{cc}
(m^2_{\tilde{e}})_{LL} &0 \\
0  &(m^2_{\tilde{e}})_{RR}\end{array}
\right),Z^E m^2_{\tilde{e}} (Z^{E})^{\dagger} =m^{2,\textup{diag}}_{\tilde{e}},\\
(m^2_{\tilde{e}})_{LL} =m_l^2+ \frac{1}{2} v_{d}^{2} |Y_{e}|^2 +\frac{1}{8}(g_1^2-g_2^2)(v_{d}^{2}- v_{u}^{2}) -g_1 v_S M_D^B-g_2v_TM_D^W ,\\
(m^2_{\tilde{e}})_{RR} = m_r^2+\frac{1}{2}v_d^2|Y_e|^2+\frac{1}{4}g_1^2( v_{u}^{2}- v_{d}^{2})+2g_1v_SM_D^B.
\end{array}
\label{sl}
\end{equation}
The sources of LFV are the off-diagonal entries of the $3\times 3$ soft supersymmetry breaking matrices $m_l^2$ and $m_r^2$ in Eqs.(\ref{sn}, \ref{sl}). From Eq.(\ref{sl}) we can see that the left-right slepton mass mixing is absent in MRSSM, whereas the $A$ terms are present in MSSM.

The mass matrix for up squarks and down squarks, and the relevant diagonalization procedure are
\begin{equation}
\begin{array}{l}
m^2_{\tilde{u}} = \left(
\begin{array}{cc}
(m^2_{\tilde{u}})_{LL} &0 \\
0  &(m^2_{\tilde{u}})_{RR}\end{array}
\right), Z^U m^2_{\tilde{u}} (Z^{U})^{\dagger} =m^{2,\textup{diag}}_{\tilde{u}}, \\
m^2_{\tilde{d}} = \left(
\begin{array}{cc}
(m^2_{\tilde{d}})_{LL} &0 \\
0  &(m^2_{\tilde{d}})_{RR}\end{array}
\right),Z^D m^2_{\tilde{d}} (Z^{D})^{\dagger} =m^{2,\textup{diag}}_{\tilde{d}},\\
(m^2_{\tilde{u}})_{LL} =m_{\tilde{q}}^2+ \frac{1}{2} v_{u}^{2} |Y_{u}|^2
+\frac{1}{24}(g_1^2-3g_2^2)(v_{u}^{2}- v_{d}^{2}) +\frac{1}{3}g_1 v_S M_D^B+g_2v_TM_D^W ,\\
(m^2_{\tilde{u}})_{RR} = m_{\tilde{u}}^2+\frac{1}{2}v_u^2|Y_u|^2+\frac{1}{6}g_1^2( v_{d}^{2}- v_{u}^{2})-\frac{4}{3} g_1v_SM_D^B,\\
(m^2_{\tilde{d}})_{LL} =m_{\tilde{q}}^2+ \frac{1}{2} v_{d}^{2} |Y_{d}|^2
+\frac{1}{24}(g_1^2+3g_2^2)(v_{u}^{2}- v_{d}^{2}) +\frac{1}{3}g_1 v_S M_D^B-g_2v_TM_D^W ,\\
(m^2_{\tilde{d}})_{RR} = m_{\tilde{d}}^2+\frac{1}{2}v_d^2|Y_d|^2+\frac{1}{12}g_1^2( v_{u}^{2}- v_{d}^{2})+\frac{2}{3} g_1v_SM_D^B.
\end{array}
\label{sud}
\end{equation}
\begin{figure}[htbp]
\setlength{\unitlength}{1mm}
\centering
\begin{minipage}[c]{1\columnwidth}
\includegraphics[width=0.85\columnwidth]{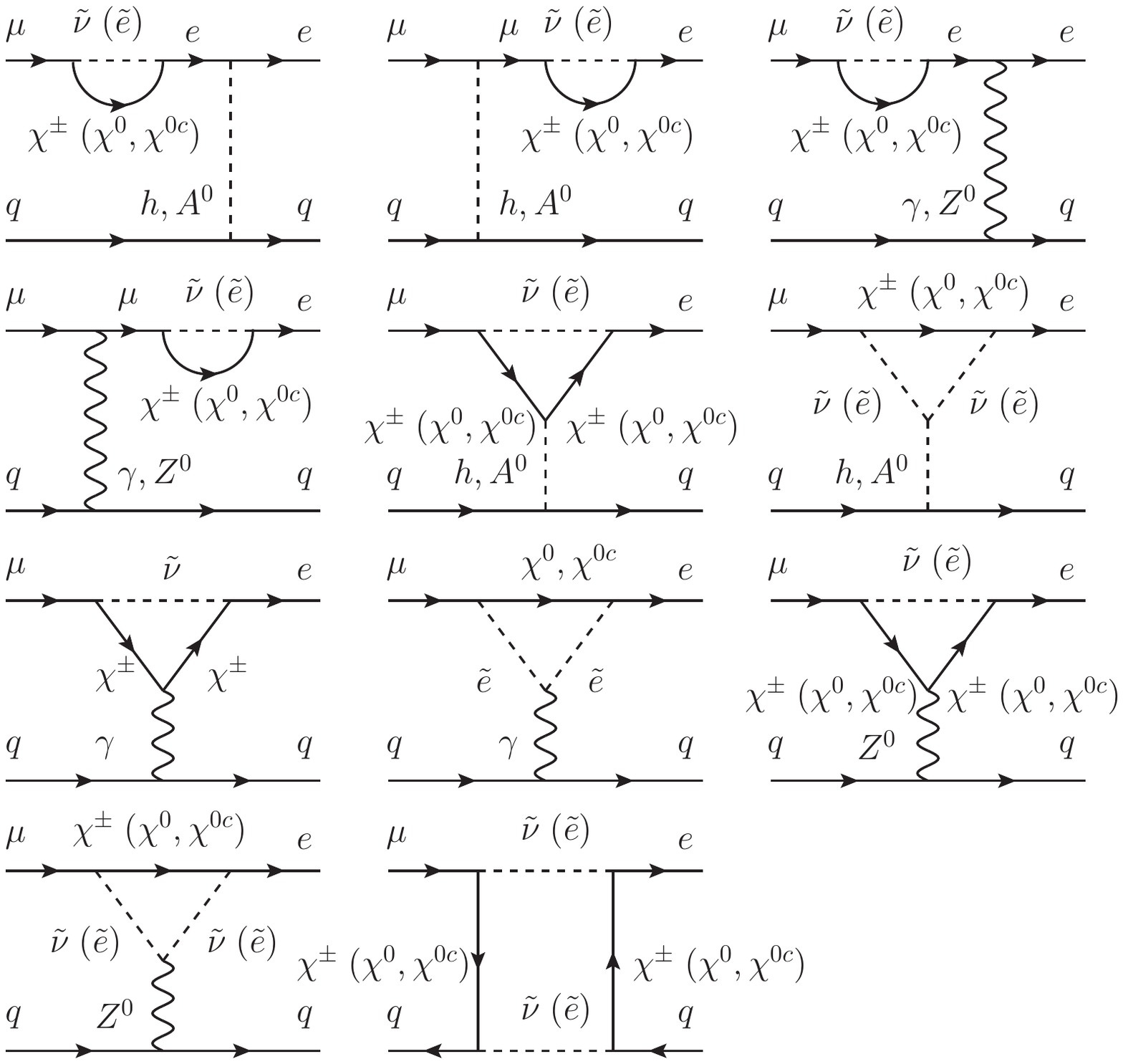}
\end{minipage}
\caption[]{One loop Feynman diagrams contributing to $\mu-e$ conversion in MRSSM.}
\label{FeynD}
\end{figure}

The MRSSM has been implemented in the Mathematica package SARAH \cite{SARAH, SARAH1, SARAH2}, and we use the Feynman rules generated with SARAH-4.14.3 in our work.
In MRSSM, violating of lepton flavor arises at the one loop level. In MRSSM, $\mu-e$ conversion is induced by the Feynman diagrams given in FIG.\ref{FeynD}. The various contributions to this process can be classified into Higgs penguins, $\gamma$ penguins, Z penguins and box diagrams. In the effective Lagrangian method, one can derive the effective Lagrangian relevant for $\mu-e$ conversion as \cite{Flavor}
\begin{eqnarray}
{\mathcal L}_{eff}&=&e\bar{l_e}\gamma^\mu\Big(K^L_1P_L+K^R_1P_R\Big)l_\mu A_\mu+\sum^{X,Y=L,R}_{K=S,V}B^K_{XY}\bar{l_e}\Gamma_KP_Xl_\mu\bar{d}\Gamma_KP_Yd\nonumber\\
&&+\sum^{X,Y=L,R}_{K=S,V}C^K_{XY}\bar{l_e}\Gamma_KP_Xl_\mu\bar{u}\Gamma_KP_Yu+h.c.
\label{eff}
\end{eqnarray}
The conversion rate $CR(\mu-e,Nucleus)$ in nuclei can be calculated by
\begin{eqnarray}
CR(\mu-e,Nucleus)&=&\sum_{X=L,R}\frac{p_eE_em_\mu^3G_F^2\alpha^3Z_{eff}^4F_p^2}{8\pi^2Z\Gamma_{capt}}\nonumber\\
&&\times\Big|(Z+N)\Big(g^{(0)}_{XV}+g^{(0)}_{XS}\Big)+(Z-N)\Big(g^{(1)}_{XV}+g^{(1)}_{XS}\Big)\Big|^2.
\label{CR}
\end{eqnarray}
Here $p_e$ and $E_e$ ($\sim m_\mu$ in the numerical evaluation) are the momentum and energy of the electron. $G_F$ and $\alpha$ are the Fermi constant and the fine structure constant, respectively. $Z_{eff}$ is the effective atomic charge. Z and N are the number of protons and neutrons in the nucleus. $F_p$ is the nuclear form factor and $\Gamma_{capt}$ is the total muon capture rate. The values of $Z_eff$, $F_p$ and $\Gamma_{capt}$ that will be used in the phenomenological analysis below are given in Table. \ref{Factors}.
At quark level, the $g^{(i)}_{XK}$ factors (with i=0,1, X=L,R and K=S,V) can be written as combinations of effective couplings
\begin{eqnarray}
g^{(i)}_{XK}&=&\frac{1}{2}\sum_{q=u,d,s}\Big(g_{XK(q)}G^{(q,p)}_K+(-1)^{i}g_{XK(q)}G^{(q,n)}_K\Big).\nonumber
\end{eqnarray}
The values of $G_K$ factors are $G^{(u,p)}_S$=$G^{(d,n)}_S$=5.1, $G^{(d,p)}_S$=$G^{(u,n)}_S$=4.3, $G^{(s,p)}_S$=$G^{(s,n)}_S$=2.5, $G^{(u,p)}_V$=$G^{(d,n)}_V$=2, $G^{(d,p)}_V$=$G^{(u,n)}_V$=1. The $g_{XK(q)}$ coefficients can be written as combinations of Wilson coefficients
\begin{eqnarray}
g_{LV(q)}=\frac{\sqrt{2}}{G_F}\Big(e^2Q_q(K^L_1-K^R_2)-\frac{1}{2}(C^{VLL}_{llqq}+C^{VLR}_{llqq})\Big),
g_{LS(q)}=-\frac{\sqrt{2}}{2G_F}\Big(C^{SLL}_{llqq}+C^{SLR}_{llqq}\Big),\nonumber
\end{eqnarray}
where $Q_q$ are the electric charge of quarks, $C^{SLL}_{llqq}$ equals $B^{K}_{XY}$ $(C^{K}_{XY})$ for d-quarks (u-quarks), $g_{RV(q)}=g_{LV(q)}|L\rightarrow R$ and $g_{RS(q)}=g_{LS(q)}|L\rightarrow R$.

\begin{table}[h]
\caption{Effective atomic charges, nuclear form factors and capture rates.}
\begin{tabular}{@{}ccccccc@{}} \toprule
Nucleus $^A_Z$N&$^{27}_{13}$Al&$^{48}_{22}$Ti&$^{80}_{38}$Sb&$^{121}_{51}$Sr&$^{197}_{79}$Au&$^{208}_{82}$Pb\\
\colrule
$Z_{eff}$&11.5&17.6&25&29&33.5&34\\
$F_p$&0.64&0.54&0.39&0.32&0.16&0.15\\
$\Gamma_{capt}\times 10^{18}$&0.464079&1.70422&4.61842&6.71711&8.59868&8.84868\\
\botrule
\end{tabular}
\label{Factors}
\end{table}

\section{Numerical Analysis\label{sec3}}
\indent\indent
We now turn to the numerical analysis of the one loop corrections to $\mu-e$ conversion in nuclei in MRSSM by using the full evaluation within the framework of SARAH-4.14.3 \cite{SARAH, SARAH1, SARAH2} and SPheno-4.0.4 \cite{SPheno1,SPheno2}.
The computation is done in a low scale version of SPheno and all free parameters are given at the SUSY scale. The experimental values of Higgs mass and $W$ boson mass can impose stringent and nontrivial constraints on the model parameters. The one loop and leading two loop corrections to the lightest (SM-like) Higgs boson in MRSSM have been computed in Ref.\cite{Die3} and several sets of benchmark points are given. These benchmark points make it possible to accommodate proper Higgs boson mass of around 125 GeV in MRSSM. The Higgs sector of the benchmark points is checked against existing experimental data using HiggsBounds and HiggsSignals and the Higgs potential of the MRSSM is checked for possible presence of deeper minima in the parameter space. There are also other restrictions. The W boson mass is found in agreement with the experimental value from combined LEP and Tevatron and low energy B meson physics observables are found agreement with measurements. All benchmark points are allowed by the fits to electroweak precision parameters S,T and U. The particle mass spectra are also shown as well as the effective couplings of the lightest Higgs particle to gauge boson and fermion pairs at leading order. A better agreement with the latest experimental value for W boson mass has been investigated in Ref.\cite{Die6}. It combines all numerically relevant contributions that are known in SM in a consistent way with all MRSSM one loop corrections. A set of updated benchmark points BMP1 is given in Ref.\cite{Die6}.

In the numerical analysis, we will use two sets of benchmark points which are taken from above references and display them in Eq.(\ref{N1}) (BMP1) and  Eq.(\ref{N2}) (BMP2). All mass parameters in Eq.(\ref{N1}) and Eq.(\ref{N2}) are in GeV or GeV$^2$.
\begin{equation}
\begin{array}{l}
\tan\beta=3,B_\mu=500^2,\lambda_d=1.0,\lambda_u=-0.8,\Lambda_d=-1.2,\Lambda_u=-1.1,
\\
M_D^B=550,M_D^W=600,\mu_d=\mu_u=500,v_S=5.9,v_T=-0.33,\\
(m^2_l)_{11}=(m^2_l)_{22}=(m^2_l)_{33}=(m^2_r)_{11}
=(m^2_r)_{22}=(m^2_r)_{33}=1000^2,\\
(m^2_{\tilde{q}})_{11}=(m^2_{\tilde{u}})_{11}=(m^2_{\tilde{d}})_{11}=(m^2_{\tilde{q}})_{22}
=(m^2_{\tilde{u}})_{22}=(m^2_{\tilde{d}})_{22}=2500^2,\\
(m^2_{\tilde{q}})_{33}=(m^2_{\tilde{u}})_{33}=(m^2_{\tilde{d}})_{33}=1000^2,m_T=3000,m_S=2000.
\end{array}\label{N1}
\end{equation}
\begin{equation}
\begin{array}{l}
\tan\beta=10,B_\mu=300^2,\lambda_d=1.1,\lambda_u=-1.1,\Lambda_d=-1.0,\Lambda_u=-1.0,
\\
M_D^B=1000,M_D^W=500,\mu_d=\mu_u=400,v_S=1.3,v_T=-0.19,\\
(m^2_l)_{11}=(m^2_l)_{22}=(m^2_l)_{33}=(m^2_r)_{11}
=(m^2_r)_{22}=(m^2_r)_{33}=1000^2,\\
(m^2_{\tilde{q}})_{11}=(m^2_{\tilde{u}})_{11}=(m^2_{\tilde{d}})_{11}=(m^2_{\tilde{q}})_{22}
=(m^2_{\tilde{u}})_{22}=(m^2_{\tilde{d}})_{22}=2500^2,\\
(m^2_{\tilde{q}})_{33}=(m^2_{\tilde{u}})_{33}=(m^2_{\tilde{d}})_{33}=1000^2,m_T=3000,m_S=2000.
\end{array}\label{N2}
\end{equation}

In following numerical analysis, the values in Eq.(\ref{N1}) and Eq.(\ref{N2}) will be used as default. Note that, the off-diagonal entries of squark mass matrices $m^2_{\tilde{q}}$, $m^2_{\tilde{u}}$, $m^2_{\tilde{d}}$ and slepton mass matrices $m^2_l$, $m^2_r$ in Eq.(\ref{N1}) and Eq.(\ref{N2}) are zero, i.e., the flavour mixing of squark and slepton is absent. Similarly to most supersymmetry models, the LFV processes in MRSSM originate from the off-diagonal entries of the soft breaking terms $m_{l}^{2}$ and $m_{r}^{2}$, which are parameterized by mass insertion
\begin{align}
(m^{2}_{l})_{IJ}=\delta ^{IJ}_{l}\sqrt{(m^{2}_{l})_{II}(m^{2}_{l})_{JJ}},
(m^{2}_{r})_{IJ}=\delta^{IJ}_{r}\sqrt{(m^{2}_{r})_{II}(m^{2}_{r})_{JJ}},\label{MI}
\end{align}
where $I,J=1,2,3$. To decrease the number of free parameters involved in our calculation, we assume that the off-diagonal entries of $m_{l}^{2}$ and $m_{r}^{2}$ in Eq.(\ref{MI}) are equal, i.e., $\delta ^{IJ}_{l}$ = $\delta ^{IJ}_{r}$ = $\delta ^{IJ}$.

The experimental limits on LFV decays, such as radiative two body decays $l_2\rightarrow l_1\gamma$, leptonic three body decays $l_2\rightarrow 3l_1$, can give strong constraints on the parameters $\delta ^{IJ}$. In the following, we will use LFV decays $\mu\rightarrow e\gamma$ to constrain the parameters $\delta ^{12}$ which has been discussed in Ref.\cite{sks2}. It is noted that $\delta ^{23}$ and $\delta ^{13}$ have been set zero in following discussion since they have no effect on the predictions of CR($\mu-e$,Nucleus). Current limits of LFV decays $\mu\rightarrow e\gamma$ is BR$(\mu\rightarrow e\gamma)<4.2\times 10^{-13}$ from MEG \cite{MEG} and new sensitivity for this decay channel in the future projects will be BR$(\mu\rightarrow e\gamma)\sim 6\times 10^{-14}$ from MEG II \cite{MEG1}.

\begin{figure}[htbp]
\setlength{\unitlength}{1mm}
\centering
\begin{minipage}[c]{1\columnwidth}
\includegraphics[width=0.4\columnwidth]{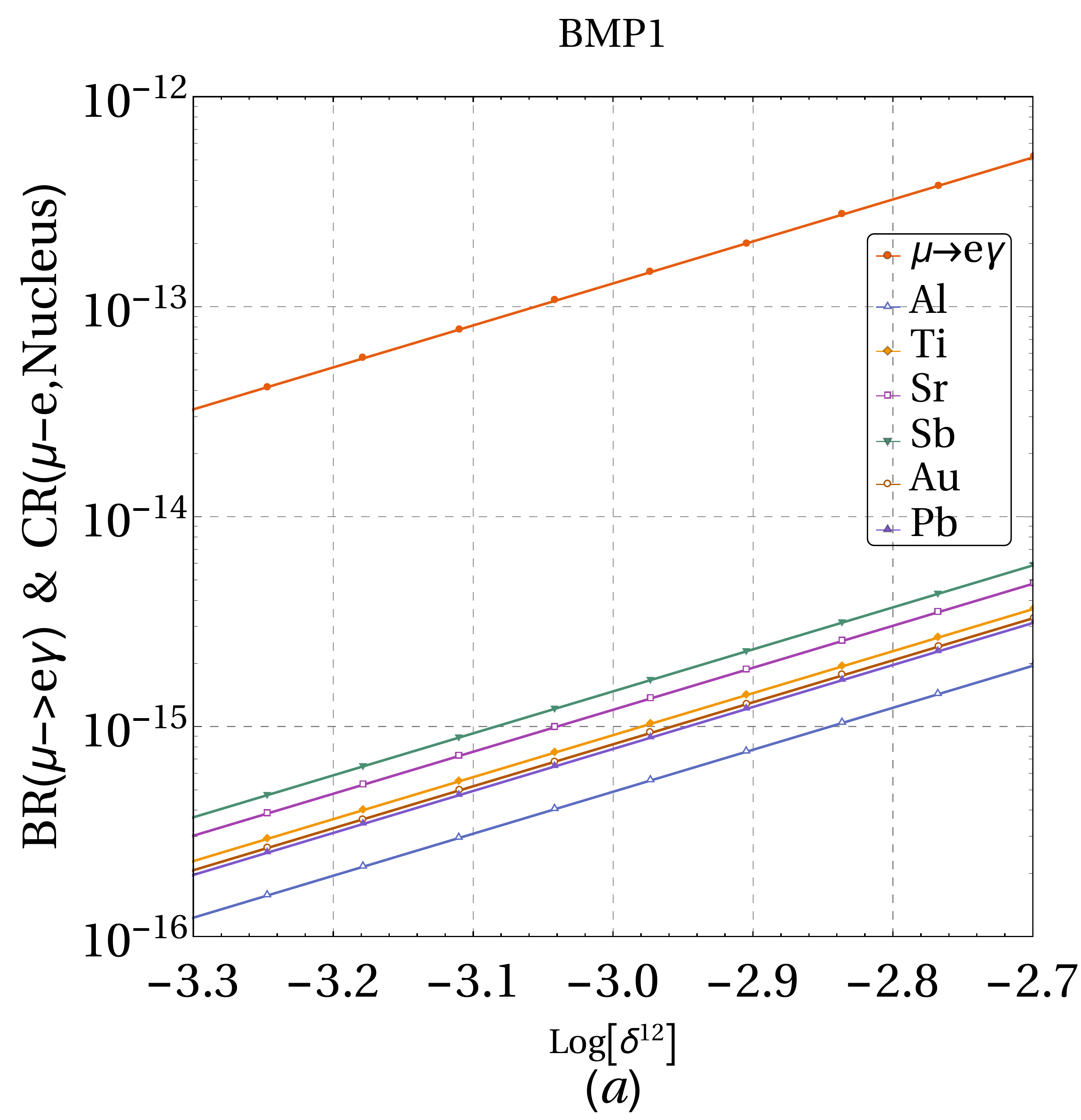}%
\includegraphics[width=0.4\columnwidth]{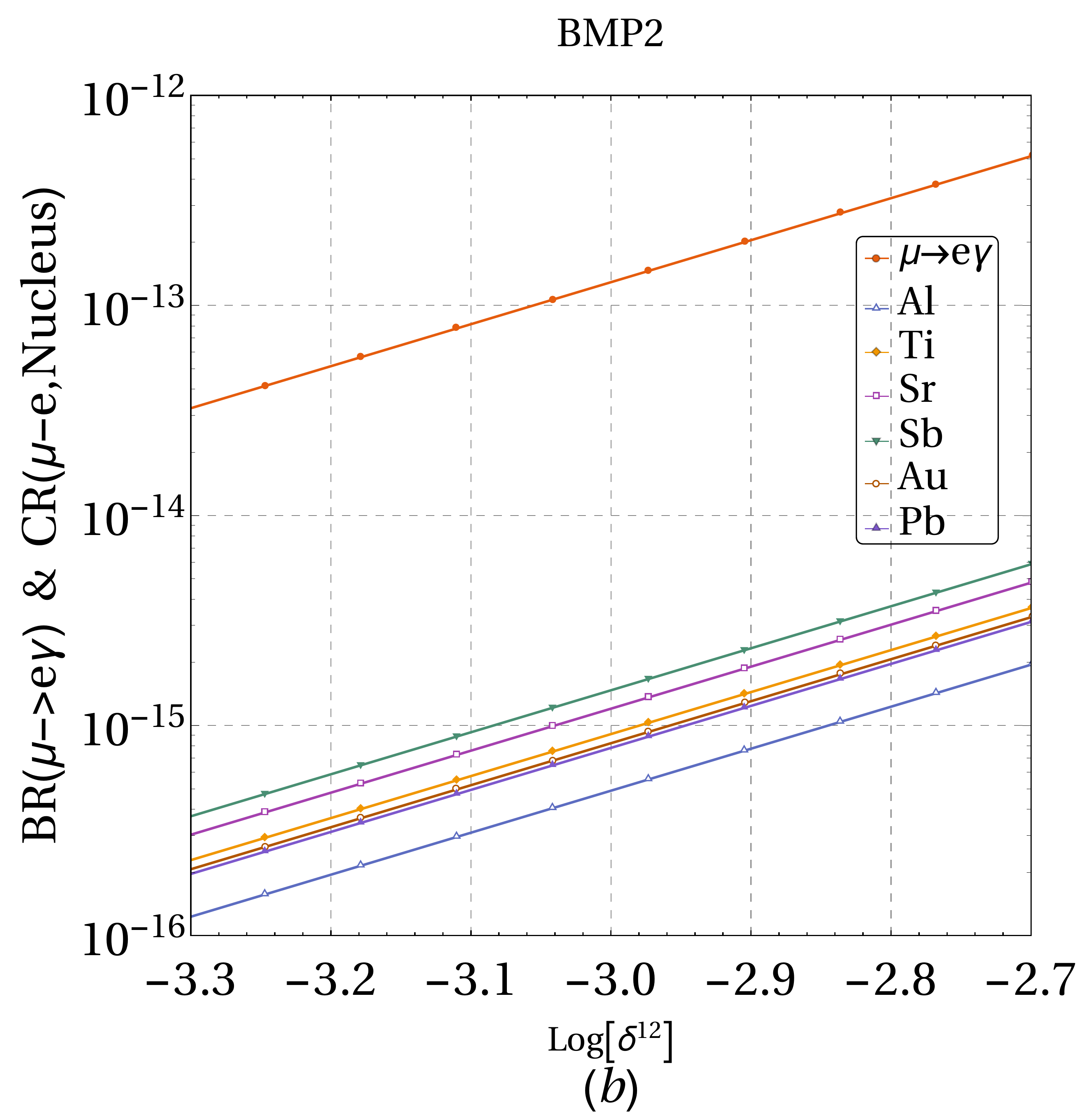}
\end{minipage}
\caption[]{Dependence of BR($\mu\rightarrow e\gamma$) and CR($\mu-e$,Nucleus) on the logarithm of mass insertion parameter $\delta^{12}$ to the base 10. All other parameters are set to the values of benchmark points BMP1 (a) and BMP2 (b).}
\label{figD12}
\end{figure}

In FIG.\ref{figD12} the predictions for BR($\mu\rightarrow e\gamma$) and CR($\mu-e$,Nucleus) for Al, Ti, Sr, Sb, Au, and Pb are shown as a function of mass insertion parameter $\delta ^{12}$ with BMP1 (a) and BMP2 (b). The prediction for BR($\mu\rightarrow e\gamma$) exceeds the future experiment sensitivity at $\delta ^{12}\sim 0.001$. In a recent Ref.\cite{sks2} the analytical computation and discussion of BR($\mu\rightarrow e\gamma$) in MRSSM has been performed. The valid region for $\delta ^{12}$ calculated in Ref.\cite{sks2} with the Mathematica package Package-X is compatible with that in this work calculated with SARAH and SPheno. We clearly see that both the predictions for BR($\mu\rightarrow e\gamma$) and CR($\mu-e$,Nucleus) in nuclei are sensitive to $\delta ^{12}$, and they increase along with the increase of $\delta ^{12}$ which have a same behavior as those in most SUSY models(e.g. \cite{sks4}). At $\delta^{12}\sim$ 0.001, the prediction on BR($\mu\rightarrow e\gamma$) is very close to the current experimental limit, and the predictions on CR($\mu-e$,Nucleus) are around $10^{-15}-10^{-16}$ which are two orders of magnitude below current experimental limits. The predicted CR($\mu-e$,Ti) is around $10^{-15}$ and this is three orders of magnitude above future experimental sensitivity \cite{PRISM}. The predicted CR($\mu-e$,Al) is around $10^{-16}$ and this is in region of the future experimental sensitivity \cite{Mu2e,COMET}. In FIG.\ref{figD12}, the predicted CR($\mu-e$,Nucleus) with BMP1 are higher than those with BMP2, and this is because of a larger tan$\beta$ and a smaller $M_D^W$ in BMP2 than those in BMP1.

As shown in FIG.\ref{figD12}, the predicted CR($\mu-e$,Nucleus) in various nuclei are very closed to each other. A lower CR($\mu-e$,Al) together with an upper CR($\mu-e$,Sb) is predicted within the same parameter space for BMP1 and BMP2, respectively. It is compatible with the result in Ref.\cite{Kitano} which indicates the CR($\mu-e$,Nucleus) increases for a light nucleus up to the atomic number Z$<$30, is largest for Z=30-60, and becomes smaller for a heavy nucleus with Z$>$60. In the following we will display the predicted CR($\mu-e$,Nucleus) in one nucleus with $\delta^{12}=0.001$ in each plot and the predicted BR($\mu\rightarrow e\gamma$) for all points in each plot satisfy the current experimental bound.

\begin{figure}[htbp]
\setlength{\unitlength}{1mm}
\centering
\begin{minipage}[c]{1\columnwidth}
\includegraphics[width=0.4\columnwidth]{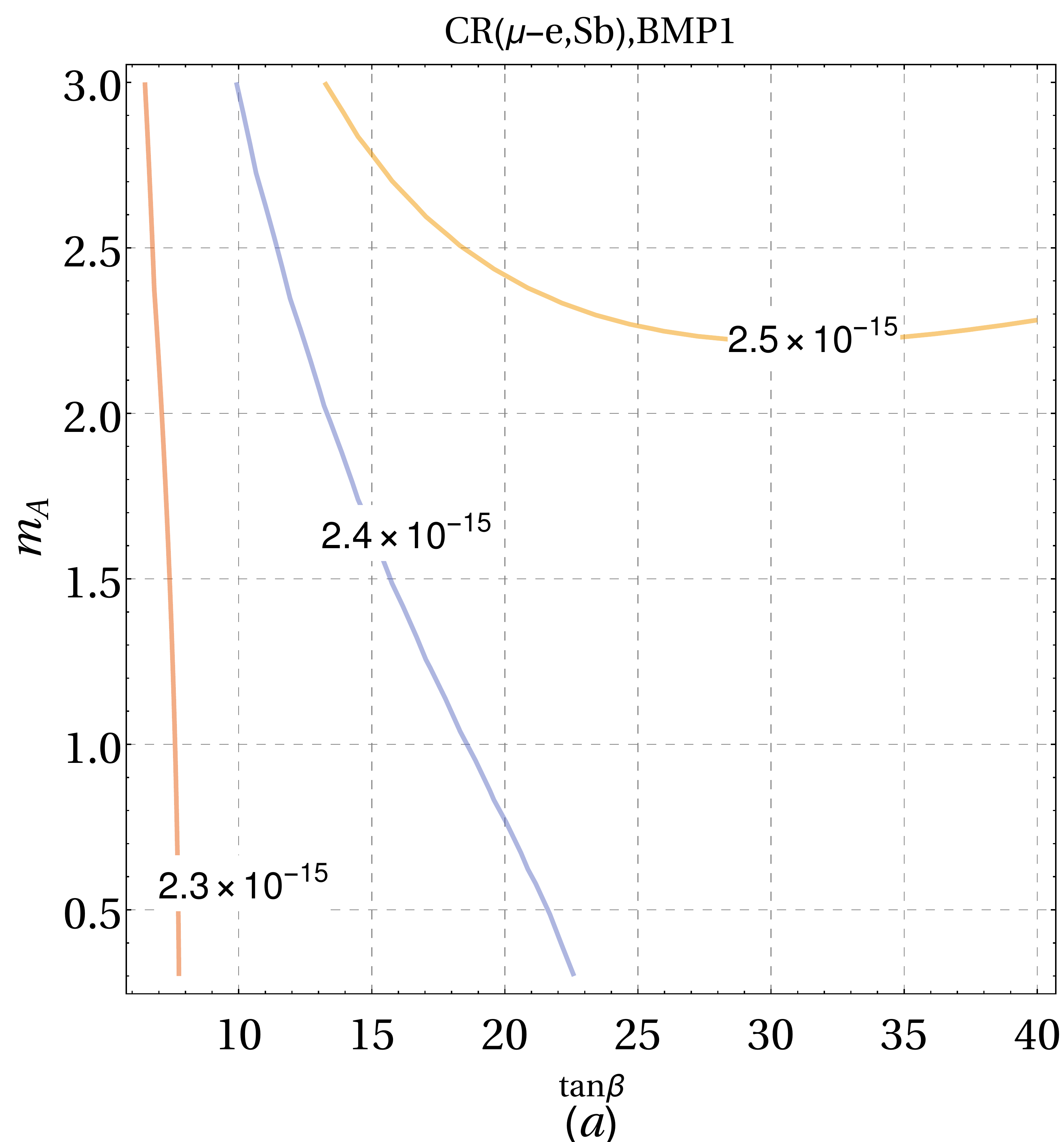}%
\includegraphics[width=0.4\columnwidth]{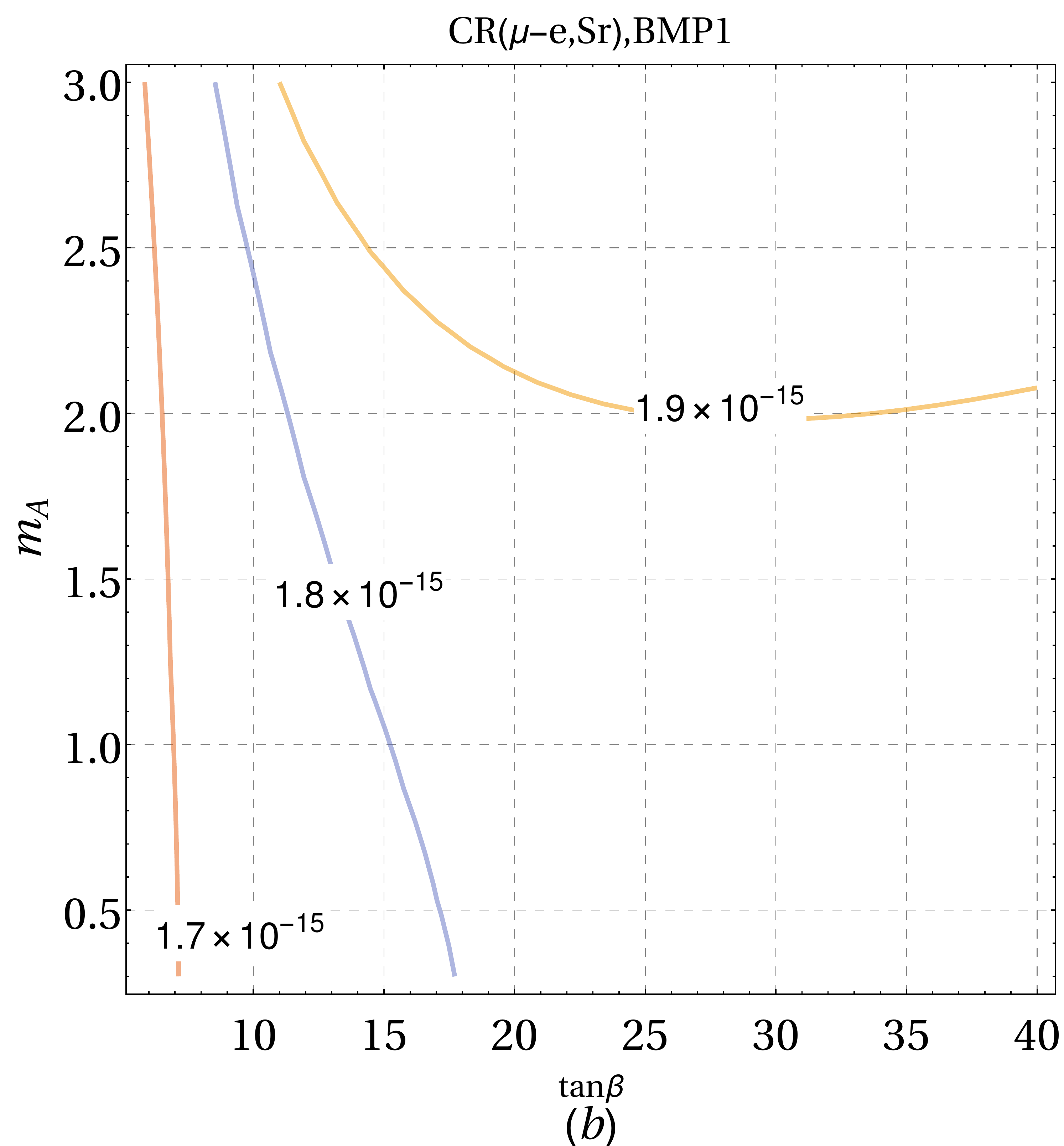}
\includegraphics[width=0.4\columnwidth]{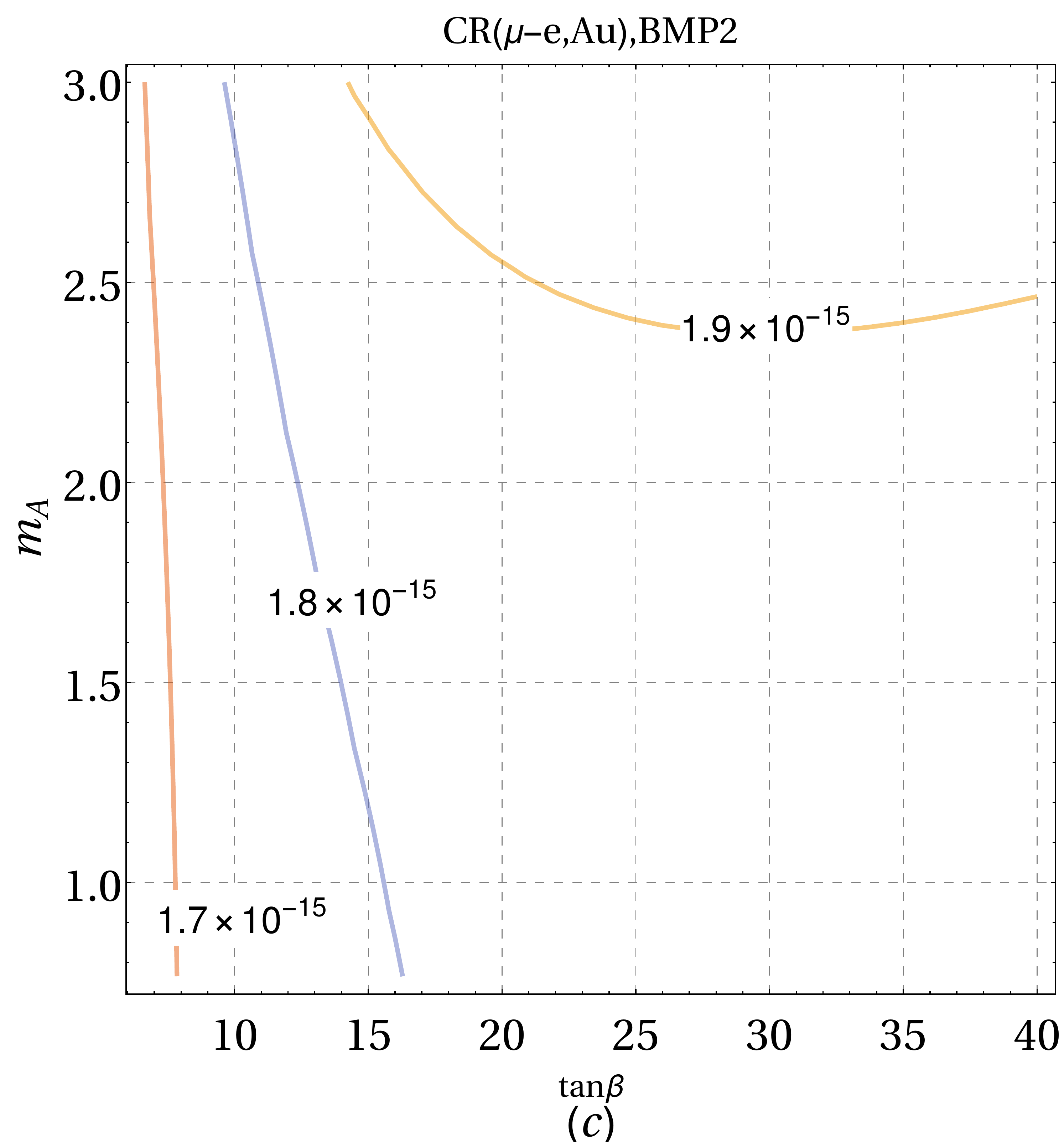}%
\includegraphics[width=0.4\columnwidth]{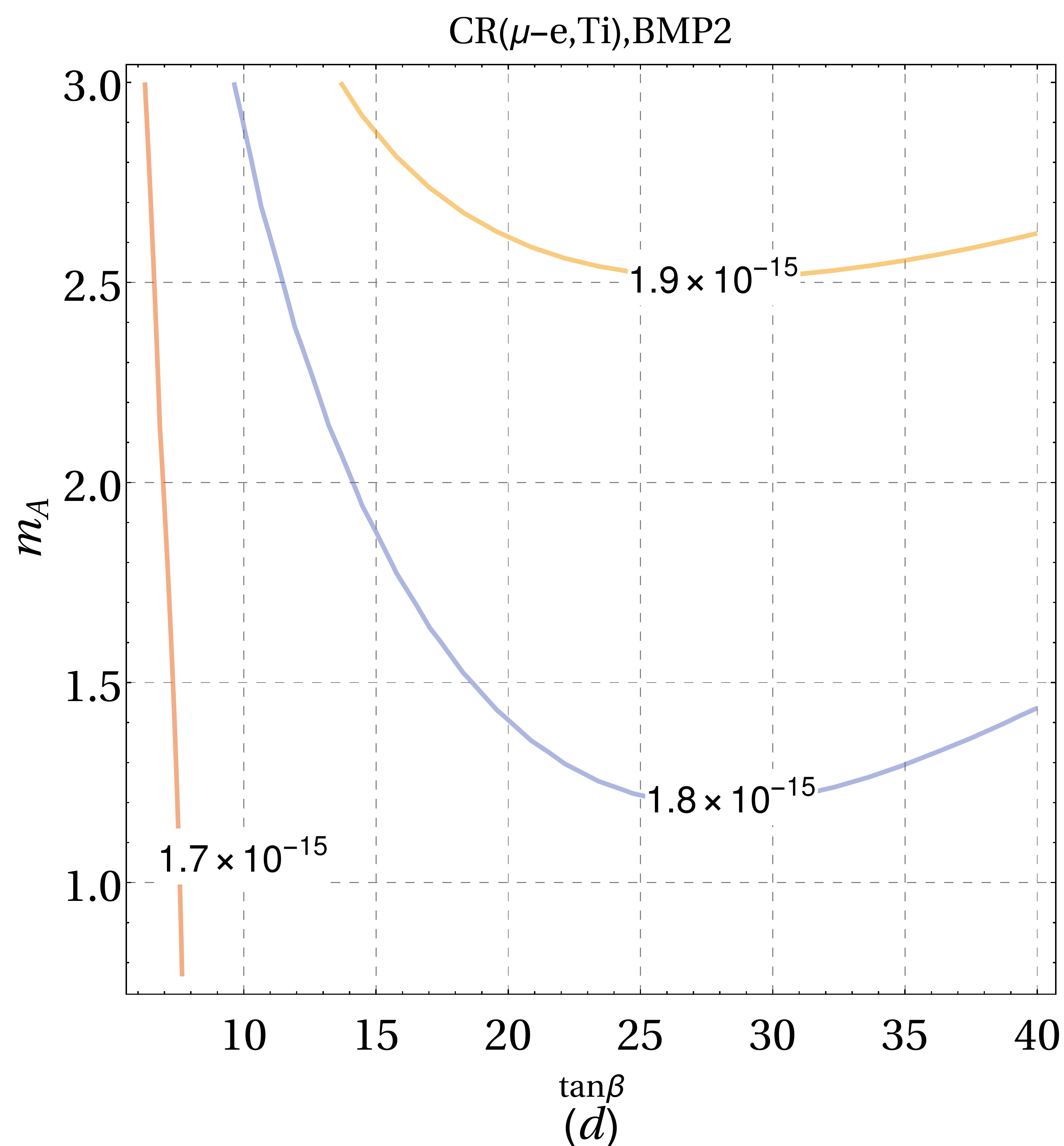}
\end{minipage}
\caption[]{Dependence of CR($\mu-e$,Nucleus) on tan$\beta$ and $m_A$. All other parameters are set to the values of benchmark points BMP1 (top row) and BMP2 (bottom row).}
\label{figBmu}
\end{figure}

In FIG.\ref{figBmu} the predictions for CR($\mu-e$,Nucleus) are shown as a function of tan$\beta$ and $m_A$. This is realized by varying parameter $B_\mu$ which is related to $m_A$ through equation $m_A^2 =\frac{2 B_\mu}{\sin2\beta}$. We clearly see that the predictions for CR($\mu-e$,Nucleus) grow as tan$\beta$ or $m_A$ grows. The predictions for CR($\mu-e$,Nucleus) in nuclei are not sensitive to tan$\beta$ or $m_A$ and take values in a narrow region. This is a striking difference to some SUSY models \cite{KSS,Guo,Dong,Zhang}. Due to the existence of the transition from $d$-Higgsino to $u$-Higgsino in MSSM, which is governed by $\mu$-term, the well-known tan$\beta$-enhancement is possible. A well-established way to understand the tan$\beta$-enhancement is provided by mass-insertion diagrams involving insertions of the $\mu$-parameter and Majorana gaugino masses. However, the $\mu$-term and Majorana gaugino masses are forbidden in MRSSM and this leads to the result that CR($\mu-e$,Nucleus) are not enhanced by tan$\beta$.

\begin{figure}[htbp]
\setlength{\unitlength}{1mm}
\centering
\begin{minipage}[c]{1\columnwidth}
\includegraphics[width=0.4\columnwidth]{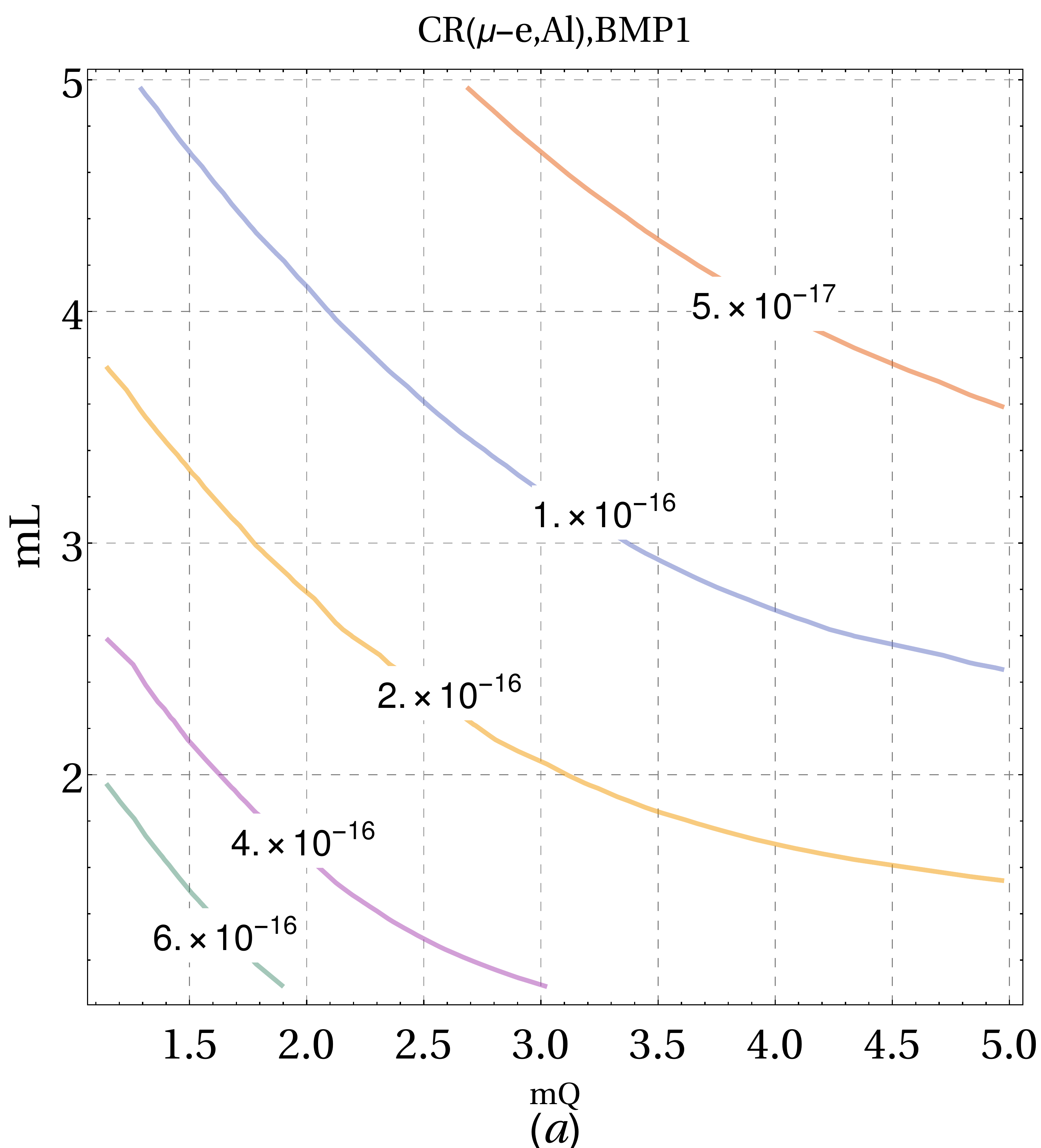}%
\includegraphics[width=0.4\columnwidth]{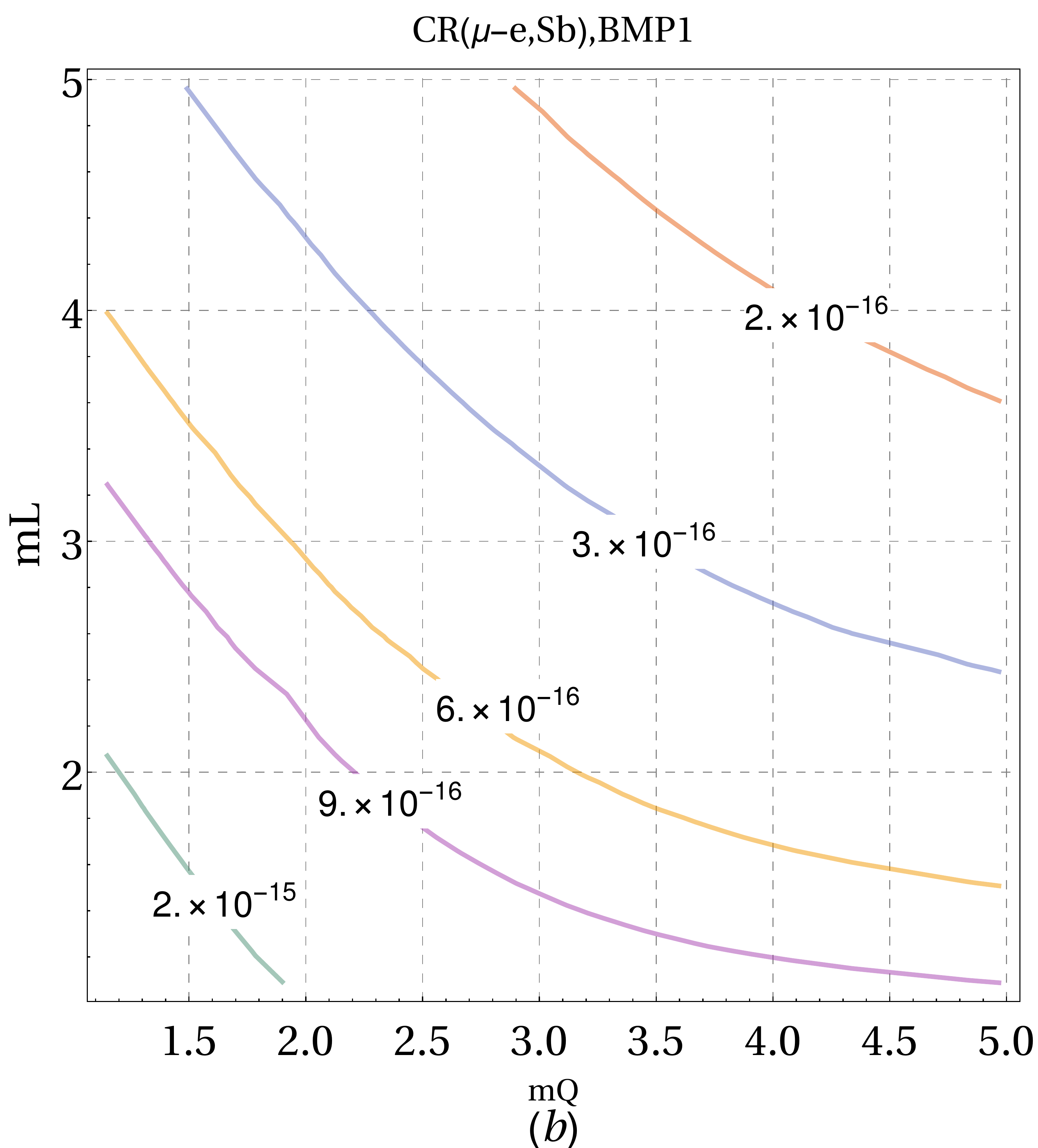}
\includegraphics[width=0.4\columnwidth]{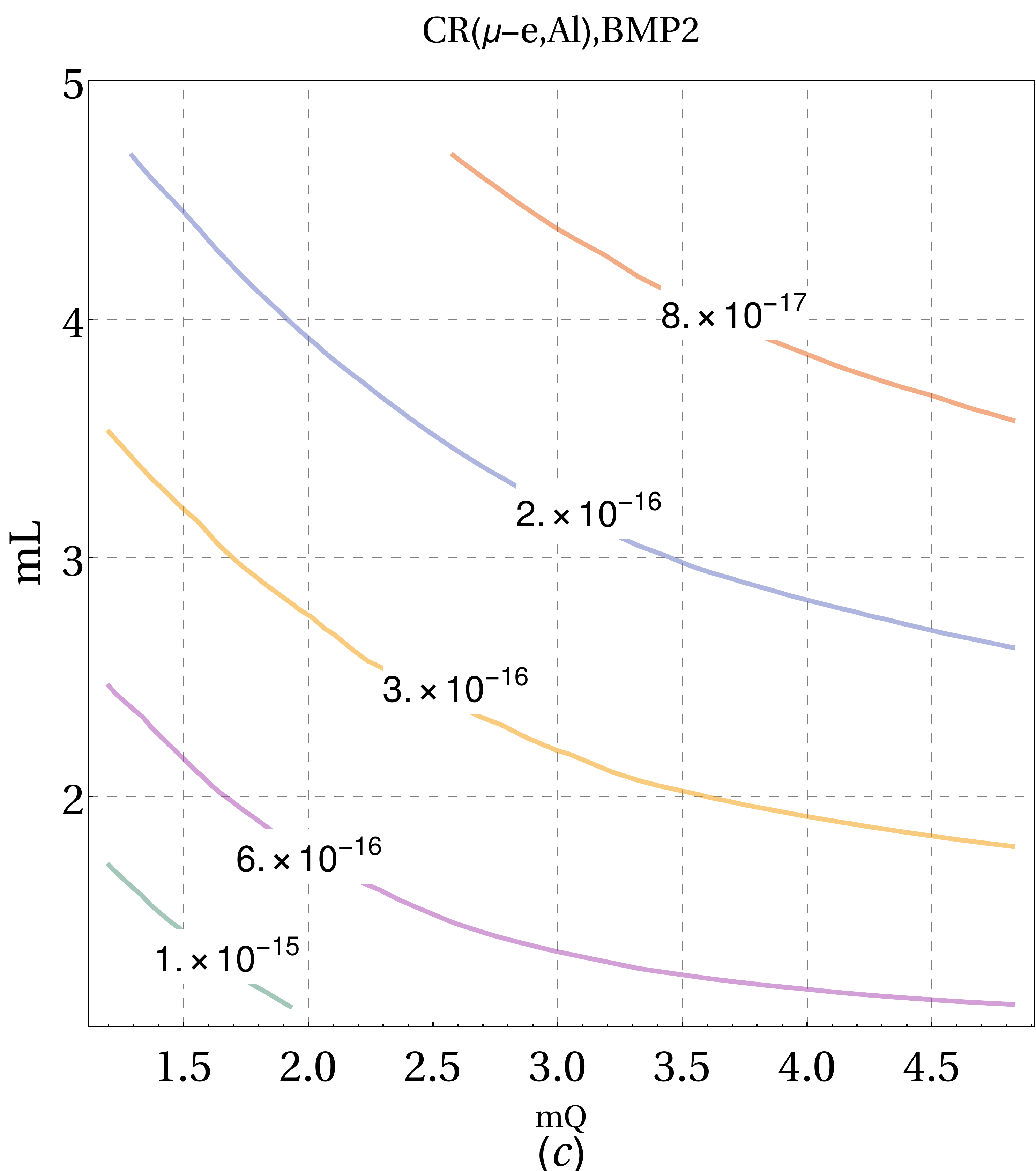}%
\includegraphics[width=0.4\columnwidth]{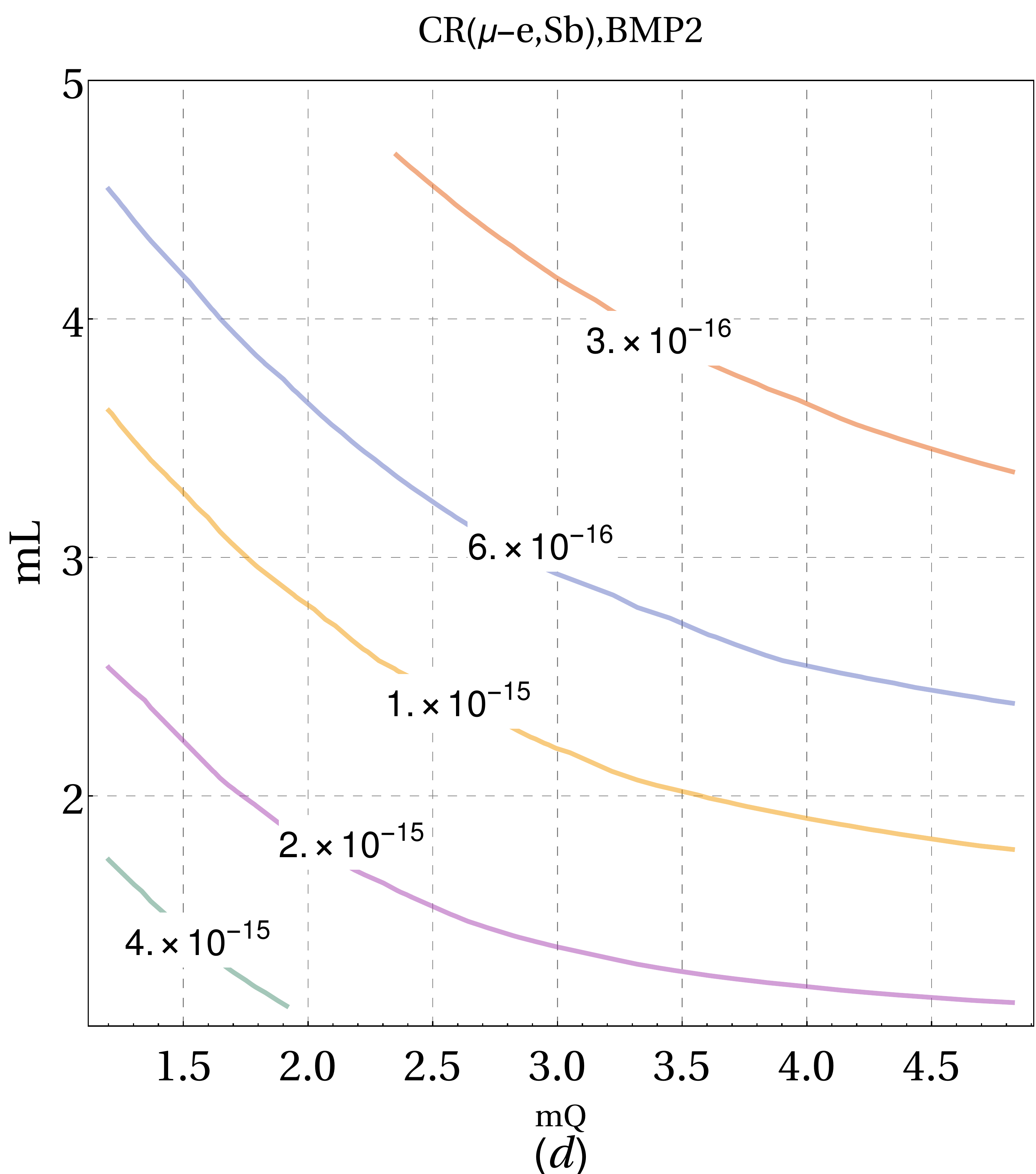}
\end{minipage}
\caption[]{Dependence of CR($\mu-e$,Nucleus) on squark mass parameter mQ and slepton mass parameter mL. All other parameters are set to the values of benchmark points BMP1 (top row) and BMP2 (bottom row), and all mass parameters are in TeV.}
\label{figMl}
\end{figure}

The predictions for CR($\mu-e$,Nucleus) in Al and Sb are shown in FIG.\ref{figMl} as a function of the squark mass parameter mQ and the slepton mass parameter mL. Here, mL=$\sqrt{(m^2_l)_{11}}=\sqrt{(m^2_l)_{22}}=\sqrt{(m^2_l)_{33}}=\sqrt{(m^2_r)_{11}}
=\sqrt{(m^2_r)_{22}}=\sqrt{(m^2_r)_{33}}$ and mQ=$\sqrt{(m^2_{\tilde{q}})_{11}}=\sqrt{(m^2_{\tilde{u}})_{11}}=\sqrt{(m^2_{\tilde{d}})_{11}}=\sqrt{(m^2_{\tilde{q}})_{22}}
=\sqrt{(m^2_{\tilde{u}})_{22}}=\sqrt{(m^2_{\tilde{d}})_{22}}=\sqrt{(m^2_{\tilde{q}})_{33}}=\sqrt{(m^2_{\tilde{u}})_{33}}=\sqrt{(m^2_{\tilde{d}})_{33}}$. We clearly see that the predictions for CR($\mu-e$,Nucleus) in nuclei are sensitive to mQ and mL, and they decrease along with the increase of mQ and mL which is described as a baseline behaviour as those in Ref.\cite{KSS,sks4}. In a wide region of mL and mQ, the predicted CR($\mu-e$,Ti) is around $10^{-16}$ and this is still two orders of magnitude above future experimental sensitivity \cite{PRISM}, and the predicted CR($\mu-e$,Al) is below $10^{-16}$ and this is in region of the future experimental sensitivity \cite{Mu2e,COMET}. Only the contribution from box diagrams for CR($\mu-e$,Nucleus) depend on the squark masses. This means the contribution from box diagrams is comparable with other diagrams.

It is noted that the predictions for CR($\mu-e$,Nucleus) in nuclei show a weak dependence on the wino-triplino mass parameter $M^W_D$, and they decrease slowly along with the increase of $M^W_D$. However, the valid region of $M^W_D$ is constrained by the boundary conditions at the unification scale, and unphysical masses of neutral Higgs and charged Higgs are obtained when $M^W_D$ above several TeV. We are also interesting to the effects from other parameters on the predictions of CR($\mu-e$,Nucleus) in MRSSM such as $M_D^B$, $\lambda_d$, $\lambda_u$, $\Lambda_d$, $\Lambda_u$,$\mu_d$ and $\mu_u$. By scanning over these parameters, the result show these parameters are also constrained in a narrow band and the predictions for CR($\mu-e$,Nucleus) take values along a narrow region.

\begin{figure}[htbp]
\setlength{\unitlength}{1mm}
\centering
\begin{minipage}[c]{1\columnwidth}
\includegraphics[width=0.4\columnwidth]{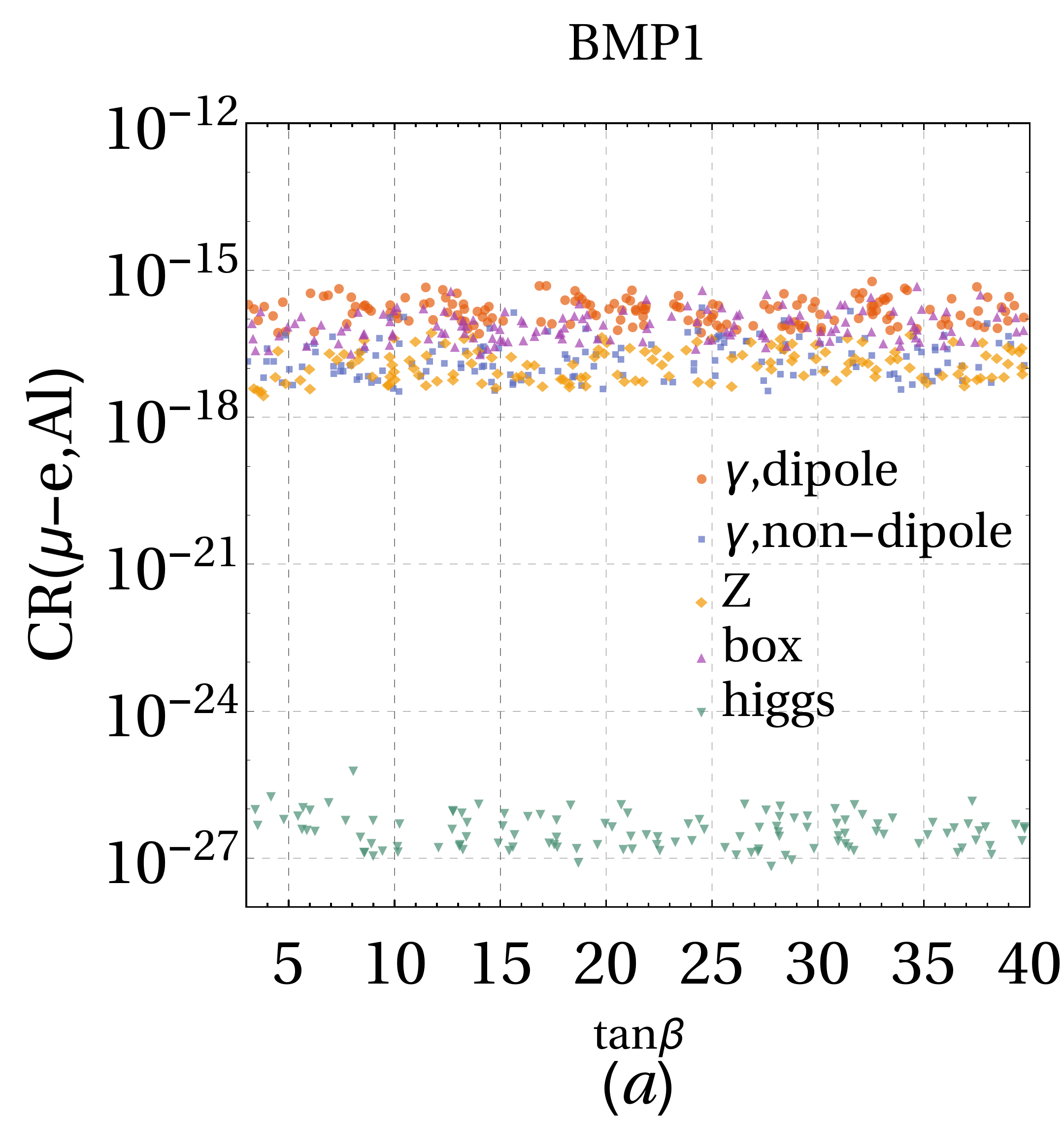}%
\includegraphics[width=0.4\columnwidth]{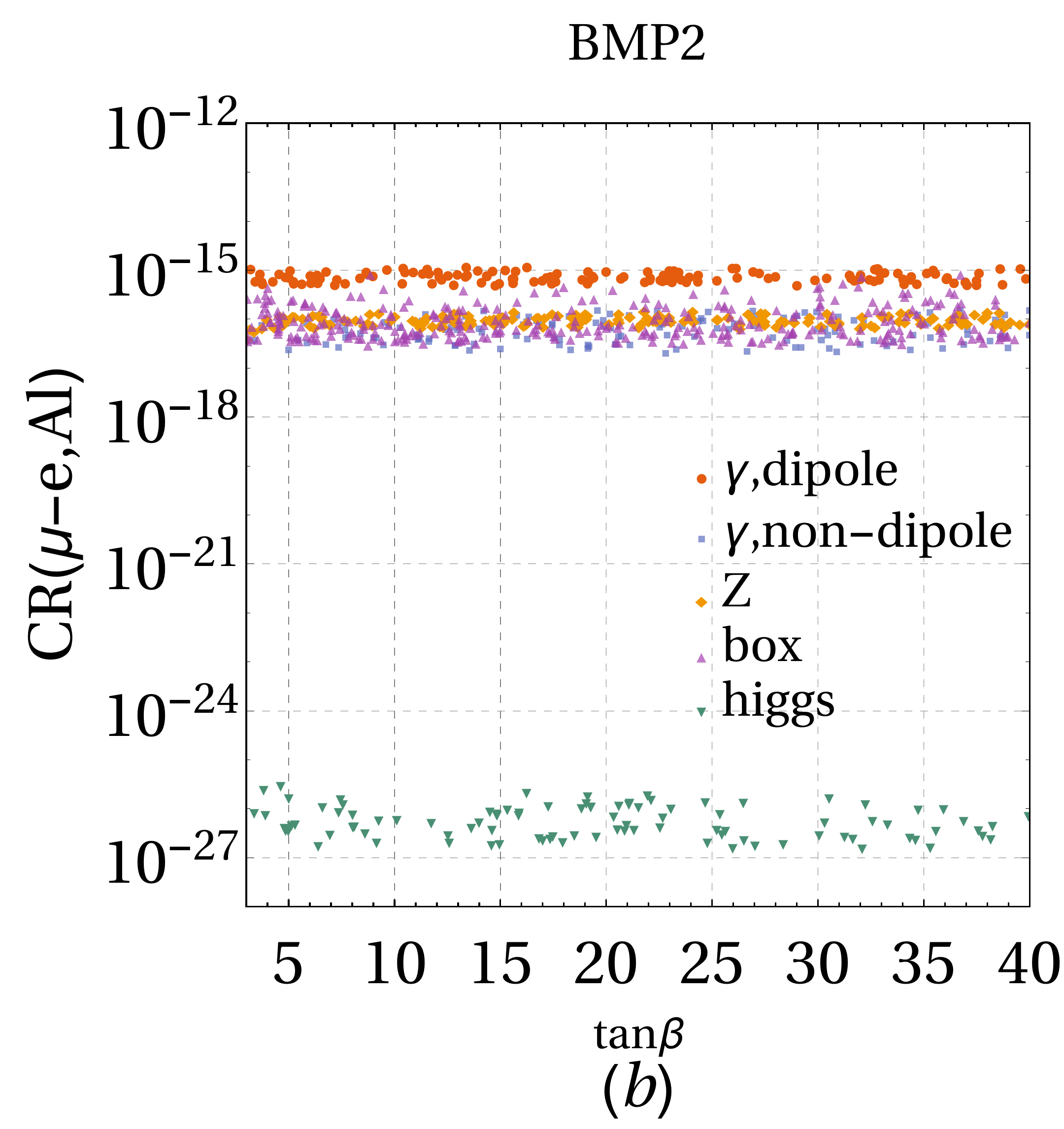}
\end{minipage}
\caption[]{Contributions from $\gamma$ dipole (blue dot), $\gamma$ non-dipole (orange square), Z penguins (green diamond), box diagrams (brown triangle) and Higgs penguins (purple inverted triangle) and to CR($\mu-e$,Al) as a function of tan$\beta$. All other parameters are set to the values of benchmark points BMP1 (a) and BMP2 (b).}
\label{figp}
\end{figure}

In FIG.\ref{figp}, we show the predictions on CR($\mu-e$,Al) as a function of tan$\beta$ with benchmark points BMP1 (a) and BMP2 (b) but independently considering the contributions from each diagram, and the values of CR($\mu-e$,Al) are given by only the listed contribution with all others set to zero. The range of input parameters for the numerical scan is given in Eq.(\ref{scan}). All other parameters are set to the values of benchmark points BMP1 (a) and BMP2 (b).
\begin{equation}
\begin{array}{l}
3<\tan\beta<40; 300 \text{ GeV} < M_D^W, M_D^B <1000 \text{ GeV};\\
9\times10^4 \text{ GeV}^2 <B_\mu<10^6 \text{ GeV}^2; 1000 \text{ GeV} < mL, mQ < 3000 \text{ GeV}.
\end{array}\label{scan}
\end{equation}

We observe that the dipole contributions $A^{L/R}_2$ from $\gamma$ penguins dominate the predictions on CR($\mu-e$,Al) similar to the case in some SUSY models (e.g.\cite{Arganda}). The contributions from Higgs penguins is negligible. In the supersymmetric seesaw model, LFV in the Higgs coupling originates from the non-holomorphic correction to the Yukawa interactions of the charged leptons \cite{Kitano1} which involves the gaugino and Higgsino mass parameters parameter. However, these parameters are absent in MRSSM. Thus the Higgs-exchange diagrams play a different rule in MRSSM from that in other SUSY models \cite{Hisano}. The predicted  CR($\mu-e$,Al) of Higgs penguins would be even smaller when $M^W_D$ close to the boundary conditions.

The non-dipole contributions $A^{L/R}_1$ from $\gamma$ penguins and the contributions from Z penguins and box diagrams are less dominant in a large parameter region. In MSSM, for a small ratio of wino mass to slepton mass, the predicted CR($\mu-e$,Nucleus) is dominated by the dipole contributions \cite{Ellis}. There is a simple relation between the CR($\mu-e$,Nucleus) and BR($\mu\rightarrow e\gamma$). Given the future experimental improvements on measuring both CR($\mu-e$, Nucleus) and BR($\mu\rightarrow e\gamma$), $\mu$-e conversion can impose limits on LFV insertions comparable to those from $\mu\rightarrow e\gamma$. The non-dipole contributions from $\gamma$/Z penguins dominate the predictions on CR($\mu-e$,Nucleus) for a small ratio of a common mass to slepton mass. In MRSSM, the predicted CR($\mu-e$,Al) of box diagrams can reach the similar magnitudes as that from the dipole contributions of $\gamma$ penguins when mQ$\sim$ 1 TeV, or the similar magnitudes as that from Z penguins or non-dipole contributions of $\gamma$ penguins when mQ $\sim$ 3 TeV. Thus, by considering the contributions form non-dipole diagrams, the predicted CR($\mu-e$,Nucleus) could be increased even larger than the predicted BR($\mu\rightarrow e\gamma$) (e.g.\cite{Ilakovac2013}). This make it possible to observe $\mu-e$ conversion in experiment while no signals of $\mu\rightarrow e\gamma$ or $\mu\rightarrow 3e$ are obtained \cite{Sato}.

\section{Conclusions\label{sec4}}
In this work, taking account of the constraints from $\mu\rightarrow e\gamma$ on the parameter space, we analyze the LFV process CR($\mu-e$,Nucleus) in the framework of the Minimal R-symmetric Supersymmetric Standard Model. In this model, R-symmetry forbids Majorana gaugino masses, $\mu$ term, $A$ terms and all left-right squark and slepton mass mixings. Due to the absent of $\mu$-term and Majorana gaugino masses, the predictions for CR($\mu-e$,Nucleus) are not enhanced by tan$\beta$. This is a main difference to MSSM.

Besides of constraints considered in Section \ref{sec3}, restrictions arising from the ATLAS and CMS searches for heavy Higgs bosons in the ditau channel should also be considered. The effect of this collider search is to impose an upper limit on tan$\beta$. The latest search for a scalar or pseudo-scalar decaying to a pair of taus with simplified exclusion likelihoods has been released by ATLAS by using 139 fb$^{-1}$ of integrated luminosity at 13 TeV \cite{ATLAS}. This model independent likelihood has been properly implemented in the new version of HiggsBounds-5 \cite{HBS}. We would like to postpone this work in our next article which analyzes the LFV decays of SM-like Higgs in MRSSM. The change of $m_A$ has a small effect on predictions of BR($\mu \rightarrow e \gamma$) and CR($\mu-e$,Nucleus) for small tan$\beta$ as shown in FIG.\ref{figBmu}. For large tan$\beta$, the effect of $m_A$ on the predictions for CR($\mu-e$,Nucleus) is also small and the latter take values along a narrow region. In Ref.\cite{ATLAS} values of tan$\beta>$ 8 and tan$\beta>$ 21 are excluded at the $95\%$ confidence level for $m_A$ = 1.0 TeV and $m_A$ = 1.5 TeV in the $M^{125}_h$ scenario of MSSM, respectively. In MRSSM, corresponding to tan$\beta$ =3 and tan$\beta$ =10, the default values of $m_A$ are $m_A$ = 0.912 TeV and $m_A$ = 0.953 TeV for BMP1 in Eq.(\ref{N1}) and BMP2 in Eq.(\ref{N2}), respectively.

In MRSSM, the theoretical predictions on CR($\mu-e$,Nucleus) mainly depend on the mass insertion $\delta^{12}$. The predictions on CR($\mu-e$,Nucleus) would be zero if $\delta^{12}$=0 is assumed. Taking account of experimental bounds on radiative decays $\mu\rightarrow e\gamma$, the values of $\delta^{12}$ is constrained around 0.001. Assuming $\delta^{12}=0.001$ and other parameter settings in Eq.(\ref{N1}), the predictions on CR($\mu-e$,Nucleus) are at the level of $\mathcal{O}(10^{-15}-10^{-16})$, which are two or three orders of magnitude above the future experimental prospects for a Al or Ti target. Thus, the LFV processes $\mu-e$ conversion in Al and Ti are very promising to be observed in near future experiment.

\begin{acknowledgments}
\indent\indent
The work has been supported partly by the National Natural Science Foundation of China (NNSFC) under Grant Nos.11905002, 11805140 and 11705045, the Scientific Research Foundation of the Higher Education Institutions of Hebei Province under Grant No. BJ2019210, the Foundation of Baoding University under Grant No. 2018Z01, the youth top-notch talent support program of the Hebei Province.
\end{acknowledgments}

\end{document}